\documentclass[12pt]{article}
\usepackage[a4paper,margin=2.2cm]{geometry}
\usepackage{rotating}
\usepackage{epsf}
\usepackage{graphics}

\title{Bounds on pressure profile and internal compactness of  static perfect fluid spheres with a positive cosmological constant}
\author{F. Shojai , A. Shojai and M. Mousavi\\
Department of Physics, University of Tehran, Tehran, Iran.} 
\date{}
\usepackage{amsmath}
\usepackage{graphicx}
\begin{document}
\maketitle
\begin{abstract}
Some theorems for a static prefect fluid sphere, i.e. a star, in the presence of a positive cosmological constant are proved. These theorems put bounds on the pressure profile and  internal compactness of the star. 
\end{abstract}
\section{Introduction}
According to Einstein's theory of gravity there are a number of constraints on the values of the characteristic functions of compact objects in the hydrostatic equilibrium state. These bounds are obtained directly from the Einstein's equations and are independent of the detailed structure and the equation of state of the stars. They are based on some acceptable assumptions: the star is spherically symmetric and static, the matter in the star is described by its non--negative mass density and the pressure and that the mean density in any arbitrary radius, does not increase outward from the center of star which is consistent with any realistic star model. In the framework of Newtonian theory of gravity, Chandrasekhar has investigated these bounds in his famous book \cite{chan}. His results include some inequalities for the central pressure, the potential energy, the mean values of gravity and pressure and so on. The general relativistic extension of these bounds is studied in \cite{matt}. The authors obtained the allowed bounds on the interior geometry, the density and pressure profile and the internal compactness, defined as $2m(r)/r$ in which $m(r)$ is the mass contained within the radius $r$, of a star. 

The most famous of these bounds is the Buchdahl--Bondi inequality \cite{bach, bond}. It is an upper bound on the total compactness of a static spherically
symmetric fluid in the form of $2M/R\le 8/9$. The modified form of this bound is derived in the presence of cosmological constant in \cite{hark1,hark2} and for a charged gravitational object in \cite{boh}.

Here we shall study how a positive cosmological constant, $\Lambda$,  effects on the other constraints of a general relativistic object which are derived in \cite{matt}.  We shall see that such a 
generalization is not trivial and requires a detailed calculation.

\section{Stellar model  in the presence of a cosmological constant}
Consider a spherically symmetric distribution of a prefect fluid  in the equilibrium state in the presence of cosmological constant. Denoting the pressure and the mass density of the fluid by $p(r)$ and $\rho(r)$ respectively. To have a realistic  star model we further assume that the average density interior to the radius r, $\overline{\rho}(r)=3m(r)/4\pi r^3$ is a non increasing function of the radial coordinate. We shall see that these assumptions lead to some inequalities for the physical quantities describing the star. 

The spherically symmetric static line element may be written in the general form:
\begin{equation}
ds^2=-f(r)dt^2+g(r)dr^2+r^2d\Omega^2
\end{equation}

Denoting the radial derivative by prime, the first equation of stellar structure is  the Newtonian definition of mass in terms of density:
\begin{equation}
m'=4\pi\rho r^2
\label{mass}
\end{equation}
The other is Tolman--Oppenheimer--Volkov equation:
\begin{equation}
p'=-\frac{\left(m+4\pi r^3 p-\frac{1}{3} \Lambda r^3\right)(\rho+p)}{r^2\left(1-\frac{2m}{r}-\frac{1}{3}\Lambda r^2\right)}
\label{tov}
\end{equation}
This equation is obtained by combining the time--time and radial--radial components of Einstein\rq{}s equations.

From the two latter equations, one can get a fundamental equation as follows:
\begin{equation}
4\pi r^2\rho=-\left[ \frac{r^2\left( (\rho+p)\left(4\pi p-\frac{1}{3}\Lambda\right)r+p'\left ( 1-\frac{1}{3}\Lambda r^2 \right ) \right)}{\rho+p-2rp'}\right ]'
\end{equation}
which for a fluid  with a given equation of state, $p=p(\rho)$, can determine the equilibrium configuration of star. 

 To obtain the complete solution, one also needs to find the geometrical structure of the space--time. This can be  easily done using the conservation of energy--momentum tensor and Einstein\rq{}s equations. The energy--momentum conservation for the perfect fluid gives:
\begin{equation}
\frac{f'}{f}=-\frac{2p'}{\rho+p}
\label{com}
\end{equation}
while the time--time component of the Einstein\rq{}s equations leads to: 
\begin{equation}
g(r)=\left( 1-\frac{2m}{r}-\frac{1}{3}\Lambda r^2 \right )^{-1}
\label{g}
\end{equation}
  
As an example consider a hypothetical star with constant density, $\rho_0 =3M/4 \pi R^3$. Then these equations are exactly solvable\cite{add}. Denoting the variables by zero index for this special solution, one has:
\begin{equation}
p_0(r)=\frac{\frac{3M}{4 \pi R^3}\left( \frac{3M}{R^3}-\Lambda \right)\left ( \sqrt{1-\left( \frac{2M}{R^3}+\frac{\Lambda}{3} \right)r^2}-\sqrt{1-\left( \frac{2M}{R^3}+\frac{\Lambda}{3} \right)R^2} \right)}{\frac{9M}{R^3}\sqrt{1-\left( \frac{2M}{R^3}+\frac{\Lambda}{3} \right)R^2}-\left( \frac{3M}{R^3}-\Lambda \right)\sqrt{1-\left( \frac{2M}{R^3}+\frac{\Lambda}{3} \right)r^2}}
\end{equation}
\begin{equation}
\sqrt{f_0(r)}=\frac{\frac{3M}{R^3}\sqrt{1-\left( \frac{2M}{R^3}+\frac{\Lambda}{3} \right)R^2}}{\left( \frac{2M}{R^3}+\frac{\Lambda}{3} \right)}-\frac{\frac{M}{R^3}-\frac{\Lambda}{3}}{\left( \frac{2M}{R^3}+\frac{\Lambda}{3} \right)}\sqrt{1-\left( \frac{2M}{R^3}+\frac{\Lambda}{3} \right)r^2}
\label{comp}
\end{equation}
The central pressure is given by:
\begin{equation}
p_0^c=\frac{\frac{3M}{4 \pi R^3}\left ( \frac{3M}{R^3}-\Lambda \right )\left( 1-\sqrt{1-\left( \frac{2M}{R^3}+\frac{\Lambda}{3} \right)R^2}\right)}{\frac{9M}{R^3}\sqrt{1-\left( \frac{2M}{R^3}+\frac{\Lambda}{3} \right)R^2}-\left( \frac{3M}{R^3}-\Lambda \right)}
\end{equation}
The central pressure becomes infinity at $M/R=2/9(1\pm \sqrt{1-3\Lambda R^2/4)}$. Thus there is some bounds on the mass to radius ratio of a uniform density star which depends on it's radius. The allowed values of internal compactness is expressed by the modified Buchdahl theorem in the presence of cosmological constant which is obtained for a general mass density in \cite{hark1,hark2} as follows:
\begin{equation}
4/9(1-\sqrt{1-3\Lambda R^2/4)}\le \frac{2M}{R}\le 4/9(1+\sqrt{1-3\Lambda R^2/4)}
\label{vay}
\end{equation}
 As a result of this  theorem a star of fixed radius which it's mass doesn't satisfy  the above bound, is not in hydrostatic equilibrium and so if it has enough mass can collapses to a deSitter--Schwarzchild black hole.
 
In this paper we shall prove some useful theorems on  different quantities describing the equilibrium configuration of a compact object.
In doing this we make the following assumptions:
\begin{itemize}
\item{I)} \textit{Einstein's general relativity is valid.}
\item{II)} \textit{There is a positive cosmological constant, $\Lambda$.}
\item{III)} \textit{The star density $\rho$ is non-negative everywhere.}
\item{IV)} \textit{The star pressure $p$ is non-negative everywhere.}
\item{V)} \textit{The star is spherically symmetric and static.}
\item{VI)} \textit{Since in the Tolman--Oppenheimer--Volkov equation the Newtonian mass appears, the average density should be defined as $\bar\rho=\frac{m(r)}{4/3\pi r^3}=\frac{4\pi\int_0^r d\tilde r\  \tilde r^2\rho(\tilde r)}{4\pi\int_0^r d\tilde r\  \tilde r^2}$. We  have to use the flat volume because in taking the average of any quantity the same measure ($4\pi r^2$) should be used in the numerator and in the denominator. For a realistic star, we assume that  $\bar\rho'\le 0$ and thus $(m/r^3)'\le 0$.\footnote{Since we assumed non-negative $\rho$ and $p$, equation (\ref{tov}) shows that, it is reasonable to assume $p$ is non-increasing on average, while the cosmological constant is not very large. In addition we suppose that the fluid allows an equation of state in the form $p=p(\rho)$ and assume that sound can propagate in the fluid, i.e. the sound velocity squared $v_s^2=\Delta p/\Delta \rho$ is non-negative. As a result,  $\bar\rho'\le 0$.}}
\end{itemize}
\section{Bounds on star properties}

Since the density and pressure are positive quantities a look at equation (\ref{tov}), we have:
\begin{equation}
p'\leq -\frac{\rho\left( m-\frac{1}{3}\Lambda r^3 \right)}{r^2\left( 1-\frac{2m}{r}-\frac{1}{3}\Lambda r^2 \right)}
\label{ine}
\end{equation}
This can be used to prove the following theorem.

\underline{\textbf{Theorem 1:}}

For any equilibrium configuration, the function
\begin{equation}
h(r)=p(r)-\frac{3}{16\pi}\frac{m}{r^3}\ln\left( 1-\frac{2m}{r}-\frac{1}{3}\Lambda r^2 \right)-\frac{3\Lambda}{8\pi}\int_0^rd\tilde{r}\frac{m}{\tilde{r}^2\left( 1-\frac{2m}{\tilde{r}}-\frac{1}{3}\Lambda \tilde{r}^2 \right)}
\label{fun}
\end{equation} 
doesn't increase outward, provided the assumptions I, II, III, IV, V and VI are valid. 

\textbf{Proof:}

Considering inequality (\ref{ine}), we find that:
\[
h'\leq -\frac{m'\left( m-\frac{1}{3}\Lambda r^3 \right)}{4\pi r^4\left( 1-\frac{2m}{r}-\frac{1}{3}\Lambda r^2 \right)}-\frac{3}{16\pi}\left(\frac{m}{r^3} \right)'\ln \left( 1-\frac{2m}{r}-\frac{1}{3}\Lambda r^2 \right)
\]
\begin{equation}
-\frac{3m}{8 \pi r^3}\frac{\left (\frac{m}{r^2}-\frac{m'}{r}-\frac{1}{3}\Lambda r\right )}{\left( 1-\frac{2m}{r}-\frac{1}{3}\Lambda r^2 \right)}-\frac{3\Lambda}{8\pi}\frac{m}{r^2\left( 1-\frac{2m}{r}-\frac{1}{3}\Lambda r^2 \right)}
\end{equation}
After some calculations and using $(m/r^3)^{\prime}\leq 0$, we obtain:
\begin{equation}
h'\leq \frac{1}{4\pi\left( 1-\frac{2m}{r}-\frac{1}{3}\Lambda r^2 \right)}\left(\frac{m}{r^3} \right)'\left ( \frac{m}{2r}+\frac{1}{3}\Lambda r^2 \right)\leq0
\end{equation}
This proves the theorem.

\hskip1cm\underline{\textbf{Corollary 1:}}
If $p_{c}$ denotes the central pressure, (\ref{fun}) leads to:
\begin{equation}
p_c\geq B_1
\end{equation}
where
\begin{equation}
B1\equiv  -\frac{3M}{16\pi R^3}\ln \left( 1-\frac{2M}{R}-\frac{1}{3}\Lambda R^2 \right)-\frac{3\Lambda}{8\pi}\int_0^Rd\tilde{r}\frac{m}{\tilde{r}^2\left( 1-\frac{2m}{\tilde{r}}-\frac{1}{3}\Lambda \tilde{r}^2 \right)}
\label{bou1}
\end{equation}
which shows a lower bound on $p_c$. 

\hskip1cm\underline{\textbf{Corollary 2:}}
There is another way to get a different lower bound on the central pressure \cite{thesis}. Integrating equation (\ref{ine}) from the center of star to it's surface where the pressure drops to zero, one obtains:
 \begin{equation}
p_c\geq \int_0^Rdr\frac{m'\left( m-\frac{1}{3}\Lambda r^3 \right)}{4\pi r^4\left( 1-\frac{2m}{r}-\frac{1}{3}\Lambda r^2 \right)}
\end{equation} 
Introducing the new variable $x(r)=2m/r +\Lambda r^2/3$, we find that:
\[
p_c\geq \frac{1}{16\pi}\int_0^R \frac{dr}{r^3(1-x)}( x+rx'-\Lambda r^2)(x-\Lambda r^2)
\]
Using the power series of $1/(1-x)$, we get:
\[
p_c\geq \frac{1}{4\pi}\int_0^Rdr \frac{1}{r^3}\left( \frac{x^2}{4r^3}-\frac{\Lambda x}{4r}+\frac{xx'}{4r^2}-\frac{\Lambda x'}{4}-\frac{\Lambda x}{2r} \right)\sum_{i=0}^\infty x^i
\]
\begin{equation}
 = \frac{1}{4\pi}\sum_{i=0}^\infty \int_0^Rdr\frac{x^i}{4r^3}\left(x-\Lambda r^2\right)^2 +\frac{1}{16\pi}\sum_{i=0}^\infty \left( \int_0^Rdr\frac{x^{1+i}x'}{r^2}-\Lambda\int_0^Rdr\ x^ix' \right)
\end{equation}
The first term is positive and the second term can be integrated by parts. This gives:
\begin{equation}
p_c\geq \frac{1}{16\pi}\left ( \left .\frac{x^{i+2}}{(i+2)r^2}\right |_0^R+\int_0^Rdr \frac{2}{r^3}\frac{x^{i+2}}{i+2}-\frac{\Lambda}{i+2}x(R)^{i+1} \right)
\end{equation}
Again the second term is positive, which shows that:
\begin{equation}
p_c\geq\frac{1}{16\pi}\sum_{i=0}^\infty \frac{x(R)^{i+2}}{(i+2)R^2}-\frac{\Lambda}{16\pi}\sum_{i=0}^{\infty}\frac{x(R)^{i+1}}{i+1}
\end{equation}
The above terms can be summed up leading to:
\begin{equation}
p_c\geq B_2
\end{equation}
where
\begin{equation}
B_2\equiv \frac{1}{16\pi}\left( \Lambda-\frac{1}{R^2}\right)\ln\left(1-\frac{2M}{R}-\frac{1}{3}\Lambda R^2 \right) +\frac{1}{16\pi R^2}\left ( \frac{2M}{R}+\frac{1}{3}\Lambda R^2\right)
\label{B2}
\end{equation}
Comparing equations (\ref{bou1}) and (\ref{bou2}) one gets:
\begin{equation}
p_c\geq \textbf{Max}\left( B_1,B_2 \right)
\label{bou2}
\end{equation}
It has to be noted that for a typical star like Sun and for the value of the cosmological constant obtained from the acceleration of the universe, $2M/R$ and $\Lambda R^2/3$ terms are very small and the $\ln$ terms can be expanded. Then it can be simply seen that $B_2$ is several order of magnitude larger than $B_1$.

\underline{\textbf{Theorem2:}}

In the presence of cosmological constant $\sqrt{f(r)}\leq \sqrt{f_0(r)}$, i.e. the time--time component of the space--time metric is less than or equal to that of the case with a constant density everywhere, provided the assumptions I, II, III, IV, V and VI are valid.

\textbf{Proof:}

Combining equations (\ref{mass}),(\ref{tov}) and (\ref{com}), it is easy to show that:
\begin{equation}
\left[ \sqrt{1-\frac{2m}{r}-\frac{1}{3}\Lambda r^2}\frac{f'}{2 f r} \right]'=\frac{\sqrt{f}}{\sqrt{1-\frac{2m}{r}-\frac{1}{3}\Lambda r^2}}\left( \frac{m}{r^3} \right )'\leq 0
\label{manfi}
\end{equation}
so that:
\begin{equation}
\frac{f'}{2 f r}\sqrt{1-\frac{2m}{r}-\frac{1}{3}\Lambda r^2}\geq \frac{1}{R}\left (  \frac{M}{R^2}-\frac{1}{3}\Lambda R \right)
\label{asl}
\end{equation}
Integrating from $r$ to $R$ leads to:
\begin{equation}
\sqrt{f(R)}-\sqrt{f(r)}\geq \frac{1}{R}\left (  \frac{M}{R^2}-\frac{1}{3}\Lambda R \right)\int_r^Rd\tilde{r} \frac{\tilde{r}}{\sqrt{1-\frac{2m}{\tilde{r}}-\frac{1}{3}\Lambda \tilde{r}^2}}
\end{equation}
Remember that the average density inside any radius $r$ is non increasing, hence:
\begin{equation}
\frac{m}{r^3}\geq \frac{M}{R^3}
\label{had}
\end{equation} 
and thus:
\begin{equation}
\sqrt{f(R)}-\sqrt{f(r)}\geq \frac{1}{R}\left (  \frac{M}{R^2}-\frac{1}{3}\Lambda R \right)\int_r^Rd\tilde{r} \frac{\tilde{r}}{\sqrt{1-\frac{2M\tilde{r}^2}{R^3}-\frac{1}{3}\Lambda \tilde{r}^2}}
\end{equation}
Performing the integration and substituting $f(R)$ from the exterior solution (i.e. Schwarzschield--deSitter solution), finally we get:
\begin{equation}
0\leq \sqrt{f(r)}\leq \frac{\frac{3M}{R^3}\sqrt{1-\frac{2M}{R}-\frac{1}{3}\Lambda R^2} -\left( \frac{M}{R^3}-\frac{1}{3}\Lambda\right) \sqrt{1-\frac{2Mr^2}{R^3}-\frac{1}{3}\Lambda r^2} }{\left(\frac{2M}{R^3}+\frac{\Lambda}{3}\right )}
\label{kisi}
\end{equation}
where the right hand side is $\sqrt{f_{0}(r)}$, (see equation (\ref{comp})), and thus
$\sqrt{f(r)}\leq \sqrt{f_{0}(r)}$.

 \hskip1cm\underline{\textbf{Corollary 1:}}
   Writing this inequality for the center of star we have $0\leq \sqrt{f(0)}\leq \sqrt{f_0(0)}$. Using the relation $0\leq \sqrt{f_0(0)}$ one gets the modified Buchdahl theorem in the presence of cosmological constant, equation (\ref{vay}), which is obtained in \cite{hark1,hark2}. 
To see this, note that squaring the relation $0\leq \sqrt{f_0(0)}$ we have:
\begin{equation}
(\Lambda R^2)^2+9\left(\frac{2M}{R}\right)\left(\frac{3}{4}\left(\frac{2M}{R}\right) -\frac{1}{3}\right)(\Lambda R^2)+ 9\left(\frac{2M}{R}\right)^2\left(\frac{9}{4}\left(\frac{2M}{R}\right) -2\right)\le 0
\end{equation}
which, using the fact that the cosmological constant is positive, is valid provided that:
\begin{equation}
\left(\frac{2M}{R}\right)^2-\frac{8}{9}\left(\frac{2M}{R}\right)+\frac{4}{27}(\Lambda R^2)\le 0
\end{equation}
which leads to the equation(\ref{vay}).

\hskip1cm\underline{\textbf{Corollary 2:}}
Using this theorem one can see that there is an upper bound on the $\overline{\rho}+3p$. The relations (\ref{tov}) and (\ref{manfi}) can be expressed respectively as:
\begin{equation}
p=\frac{1}{4\pi r}\left ( 1- \frac{2m}{r}-\frac{1}{3}\Lambda r^2\right ) \frac{f'}{2f}-\frac{\bar{\rho}}{3}+ \frac{\Lambda}{12\pi}
\label{1}
\end{equation} 
and
\begin{equation}
\frac{f'}{2f}\geq \frac{\left ( \frac{4\pi}{3}\rho_0-\frac{1}{3}\Lambda\right )r}{\sqrt{f}\sqrt{1- \frac{2m}{r}-\frac{1}{3}\Lambda r^2}}
\label{2}
\end{equation}
Using equation (\ref{1}), one can eliminate $f^{\prime}/f$ in the relation (\ref{2}) to obtain the following result:
\begin{equation}
3p+\bar{\rho}\geq \sqrt{1- \frac{2m}{r}-\frac{1}{3}\Lambda r^2}\left ( \frac{\rho_0-\frac{1}{4\pi}\Lambda}{\sqrt{f}}\right )+\frac{1}{4\pi}\Lambda
\end{equation} 

\hskip1cm\underline{\textbf{Corollary 3:}}
Another result is the existence of a lower limit on the pressure of a star at any arbitrary radius. Integrating equation (\ref{com}) from $r$ to $R$ and using equation (\ref{asl}) one has:
\begin{equation}
p(r)\sqrt{f(r)}\geq \left( \frac{M}{R^3}-\frac{1}{3}\Lambda\right) \int_r^R d\tilde{r}\frac{\tilde{r}\rho(\tilde{r})}{\sqrt{1- \frac{2m}{\tilde{r}}-\frac{1}{3}\Lambda \tilde{r}^2}}
\end{equation}
which can be written in terms of $(m/r)'$ and $\rho_0$ as follows:
\begin{equation}
p(r)\sqrt{f(r)}\geq \frac{1}{3}\left( \rho_0-\frac{1}{4\pi}\Lambda\right ) \int_r^R d\tilde{r}\frac{\left(\frac{m}{\tilde{r}}\right)'+\frac{m}{\tilde{r}^2}}{\sqrt{1- \frac{2m}{\tilde{r}}-\frac{1}{3}\Lambda \tilde{r}^2}}
\end{equation}
By adding and subtracting $\Lambda\tilde{r}/3$ in the numerator of integrand, the above integral can be decomposed into two terms. The first term is integrable and the second term, because of equation (\ref{had}) has  an upper limit. Thus:
\begin{equation}
p(r)\sqrt{f(r)}\geq -\frac{1}{3}\left( \rho_0-\frac{1}{4\pi}\Lambda\right )\left.\sqrt{1- \frac{2m}{\tilde{r}}-\frac{1}{3}\Lambda \tilde{r}^2}\right|_r^R+ \left( \frac{M}{R^3}-\frac{1}{3}\Lambda\right) \int_r^R d\tilde{r}\frac{\tilde{r}}{\sqrt{1- \frac{2M\tilde{r}^2}{R^3}-\frac{1}{3}\Lambda \tilde{r}^2}}
\end{equation}
and after a simple integration and noting that $f(r)\le f_0(r)$ and that $m/r^3$  is decreasing we have:
\[
p(r)\sqrt{f_0(r)}\geq \frac{1}{3}\left( \rho_0-\frac{1}{4\pi}\Lambda\right )\left[ \sqrt{1- \frac{2Mr^2}{R^3}-\frac{1}{3}\Lambda r^2}+\right.
\]
\begin{equation}
\left.\frac{\frac{M}{R^3}-\frac{\Lambda}{3}}{\left(\frac{2M}{R^3}+\frac{\Lambda}{3}\right)} \sqrt{1- \frac{2Mr^2}{R^3}-\frac{1}{3}\Lambda r^2}- \frac{\frac{3M}{R^3}}{\left(\frac{2M}{R^3}+\frac{\Lambda}{3}\right)} \sqrt{1- \frac{2M}{R}-\frac{1}{3}\Lambda R^2}\right]
\label{mordamazinhameformul}
\end{equation}

\underline{\textbf{Theorem3:}}

For a stellar object, one has:
\begin{equation}
\frac{m}{r}\leq -\frac{\Lambda}{6}r^2+\frac{1}{9}\left [ 
1-\frac{1+6\pi r^2 p}{2}+\frac{3}{4}\Lambda r^2+\frac{\sqrt{1+6\pi pr^2-\frac{3}{4}\Lambda r^2}}{2} \right ]
\label{compact}
\end{equation}
provided the assumptions I, II, III, IV, V and VI are valid.

\textbf{Proof:}

According to (\ref{manfi}):
\begin{equation}
\frac{f'(\tilde{r})}{2 \tilde{r} f(\tilde{r})}\sqrt{1-\frac{2m}{\tilde{r}}-\frac{1}{3}\Lambda r^2}\geq \frac{f'(r)}{2 r f(r)}\sqrt{1-\frac{2m}{r}-\frac{1}{3}\Lambda r^2}
\end{equation}
if $\tilde{r}<r$. By substituting equations (\ref{com}) and (\ref{tov}) in the right hand side  and integrating from $\tilde r=0$ to $\tilde r=r$ we get:
\begin{equation}
\sqrt{f(r)}-\sqrt{f(0)}\geq \sqrt{f(r)}S(r)\int_0^r\frac{\tilde{r}d\tilde{r}}{\sqrt{1-\frac{2m(\tilde{r})}{\tilde{r}}-\frac{\Lambda}{3}\tilde{r}^2}}
\end{equation}
where:
\begin{equation}
S(r)=\frac{m(r)+4\pi r^3p-\frac{1}{3}\Lambda r^3}{r^3\sqrt{1-\frac{2m}{r}-\frac{\Lambda}{3}r^2}}
\end{equation}
Since $m(\tilde{r})\geq m(r) \frac{\tilde{r}^3}{r^3}$:
\begin{equation}
\sqrt{f(r)}-\sqrt{f(0)}\geq \sqrt{f(r)}S(r)\int_0^r\frac{\tilde{r}d\tilde{r}}{\sqrt{1-\left ( \frac{2 m(r)}{r^3}+\frac{\Lambda}{3}\right)\tilde{r}^2}}
\end{equation}
Evaluating the right hand side integral and applying the condition that $f(0)\geq 0$, one finds:
\begin{equation}
1+\frac{m+4\pi r^3p-\frac{1}{3}\Lambda}{\left(2m+\frac{1}{3}\Lambda r^3\right)\sqrt{1-\frac{2m}{r}-\frac{\Lambda}{3}r^2}}\left( \sqrt{1-\left(\frac{2m}{r^3}-\frac{\Lambda}{3}\right)r^2}-1 \right)\geq 0
\label{it}
\end{equation}
which can be written as: 
\begin{equation}
\left ( 1-\frac{2m}{r}-\frac{\Lambda}{3}r^2 \right) \left( \frac{3m}{r}+4\pi pr^2\right)^2\geq \left(  \frac{m}{r}+4\pi r^2p-\frac{1}{3}\Lambda \right)^2
\label{wald}
\end{equation}
This expression can be simplified by defining:
\begin{equation}
\left\{
\begin{tabular}{l}
$m=\hat{m}-\frac{\Lambda}{6}r^3$\\
$4\pi p=4\pi\hat{p}+\frac{1}{2}\Lambda$\\
\end{tabular}
\right.
\label{def}
\end{equation}
In terms of these variables, equation (\ref{wald}) gives:
\begin{equation}
\left ( 1-\frac{2\hat{m}}{r}\right) \left( \frac{3\hat{m}}{r}+4\pi \hat{p}r^2\right)^2\geq \left(  \frac{\hat{m}}{r}+4\pi \hat{p}r^2 \right)^2
\end{equation}
which is exactly the same relation for $m$ and $p$  in the absence of
$\Lambda$ \cite{wald} and so it can be written as:
\begin{equation}
\frac{\hat{m}}{r}\leq\frac{2}{9}\left [ 1-6\pi\hat{p}r^2+\sqrt{1+6\pi\hat{p}r^2} \right ]
\end{equation}
Writing the above relation in terms of $m$ and $p$  we get our desired result, equation(\ref{compact}).
It represents an upper bound on the internal compactness of a star in the presence of a  cosmological constant.  To understand  this inequality better,  consider a realistic model for a star in which there is a core represented by it's 
mass $\tilde{m}$ and radius $\tilde{r}$. Outside the  core, assume that the density is below a given 
density $\tilde{\rho}$ and the equation of state is known only at this region. According to the above 
theorem, the total mass of the star is bound. To see this note that the lower bound of $\tilde{m}$ is $(4/3) \pi\tilde{r}^3\tilde{\rho}$ and (\ref{compact}) shows that it's upper bound is given by:
\begin{equation}
\tilde{m}\leq -\frac{\Lambda}{6}\tilde{r}^3+\frac{\tilde{r}}{9}\left [ 
1-\frac{1+6\pi \tilde{r}^2 \tilde{p}}{2}+\frac{3}{4}\Lambda \tilde{r}^2+\frac{\sqrt{1+6\pi \tilde{p}\tilde{r}^2-\frac{3}{4} \Lambda \tilde{r}^2}}{2} \right ]
\end{equation}
where $\tilde{p}$ is the pressure at the core boundary. Since the values of $\tilde{m}$ and $\tilde{r}$ are restricted, the total mass which is a continuous function, $M(\tilde{m},\tilde{r})$, is also bound.

Moreover the relation (\ref{it}) implies that the pressure at any radius has a maximum value which depends on the mass contained within that radius. One can see this, by multiplying equation (\ref{it}) by $1+\sqrt{1-\left( \frac{2m}{r^3}+\frac{\Lambda}{3} \right)r^2}$ and doing some simplifications:
\begin{equation}
p(r)\leq\frac{1}{4\pi r^2}\left [  1-\frac{3m}{r}+ \sqrt{1-\left( \frac{2m}{r^3}+\frac{\Lambda}{3} \right)r^2} \right ]
\end{equation}  
\section{Conclusion}
The theorems proved for a static spherically symmetric stars in the presence of a positive cosmological constant, enabled us to obtain upper and lower bounds for some of the physical properties of the star. These inequalities include upper bounds for the central pressure, the pressure profile and also for the internal compactness.

\begin{figure}[htp]
\centering
\includegraphics[scale=0.6]{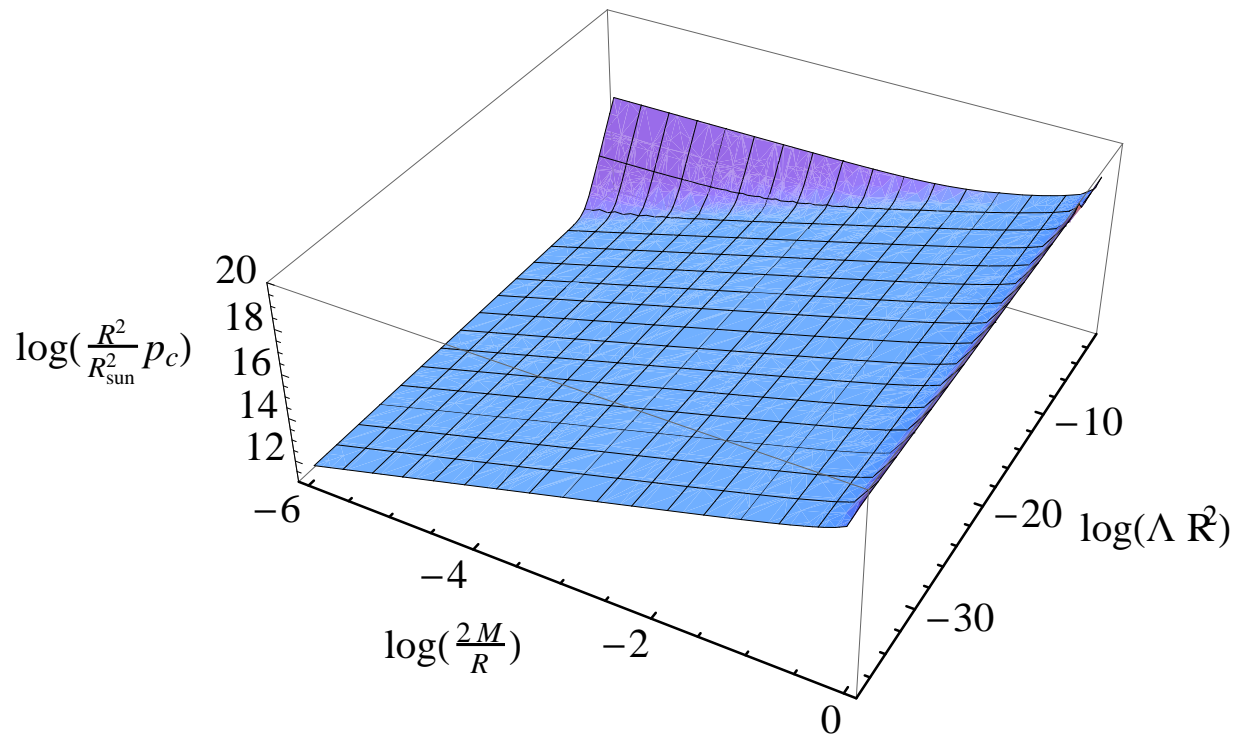}
\caption{The lower bound $B_2$ in terms of the internal compactness and the cosmological constant. The acceptable region is the upper region of the surface. $p_c$ is in atmospheres and $R_{sun}$ is used to normalize the star radius.}
\label{f1}
\end{figure}

To see how the presence of a positive cosmological constant changes these bounds, let us to investigate the $B_2$ bound on the central pressure defined by equation (\ref{B2}). The dependence of the central pressure on the internal compactness and the cosmological constant is plotted in Figure (\ref{f1}).

As it is clear from this plot, the bound has changed by a considerable value, only for very large cosmological constant.  

\begin{figure}[htp]
\centering
\includegraphics[scale=0.6]{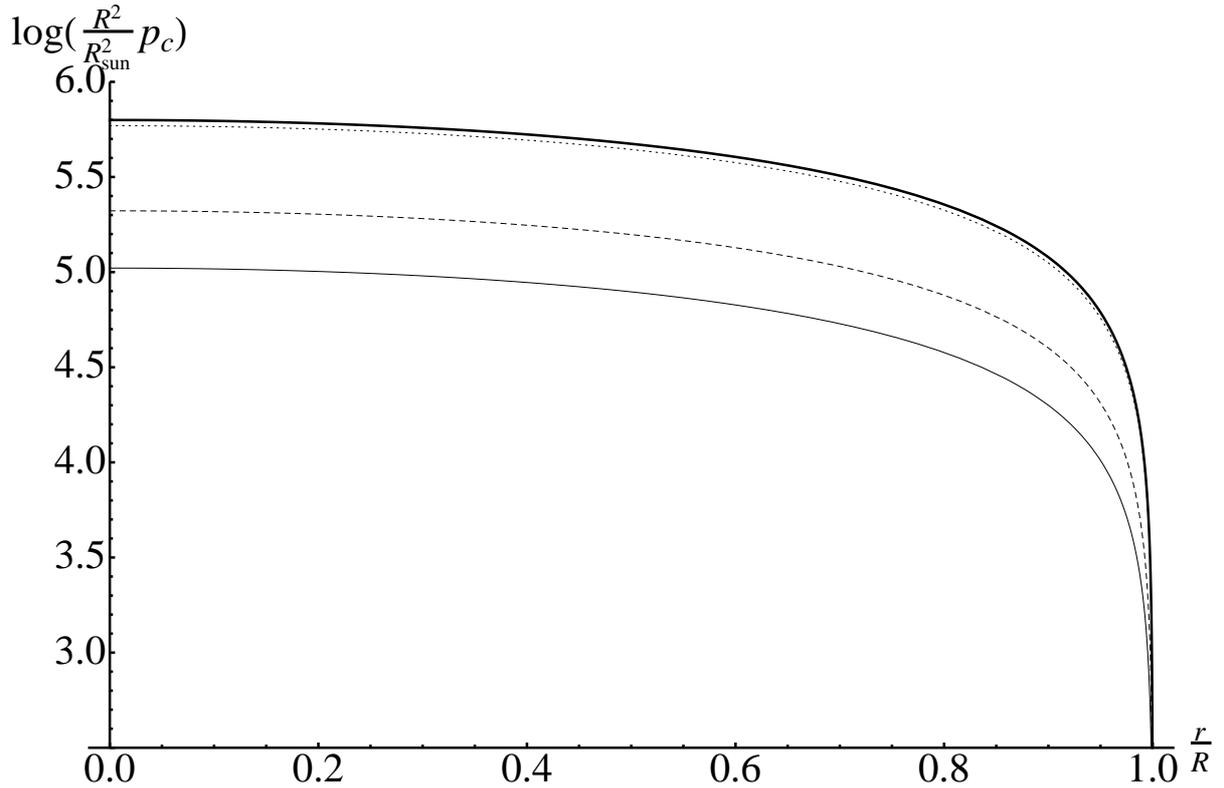}
\caption{The lower bound on the pressure profile, for different values of the cosmological constant. Here we assumed that $2M/R = 2\times 10^{-6}$. The thick, dotted, dashed and tiny lines corresponds to $\Lambda R^2 = 2\times 10^{-34}$, $\Lambda R^2= 2\times 10^{-7}$, $\Lambda R^2 = 2\times 10^{-6}$, and $\Lambda R^2 = 2.5 \times 10^{-6}$. $p_c$ is in atmospheres and $R_{sun}$ is used to normalize the star radius.}
\label{f2}
\end{figure}

 As another example consider the bound on the pressure profile given by equation (\ref{mordamazinhameformul}).
 In Figure (\ref{f2}), the lower bound of the pressure profile is plotted for different values of the cosmological constant. Again it is seen that only for very large cosmological constant the difference is considerable.

\textbf{Acknowledgment} This work is  supported by a grant from university of Tehran.

\end{document}